\documentclass{ws-ijmpa}

\def\beq{\begin{equation}}
\def\eeq{\end{equation}}
\def\C{{\cal C}}

\makeatletter
\@ifundefined{lesssim}{\def\lesssim{\mathrel{\mathpalette\vereq<}}}{}
\@ifundefined{gtrsim}{\def\gtrsim{\mathrel{\mathpalette\vereq>}}}{}
\def\vereq#1#2{\lower3pt\vbox{\baselineskip1.5pt \lineskip1.5pt
\ialign{$\m@th#1\hfill##\hfil$\crcr#2\crcr\sim\crcr}}}
\makeatother

\begin{document}

\markboth{David Langlois}
{Gravitational and cosmological properties  of  a brane-universe}

\catchline{}{}{}

\title{GRAVITATIONAL AND COSMOLOGICAL PROPERTIES OF A BRANE-UNIVERSE}

\author{\footnotesize DAVID LANGLOIS\footnote{langlois@iap.fr}}

\address{Institut d'Astrophysique de Paris, \\
98bis Boulevard Arago, 75014 Paris, France
}

\maketitle

\begin{abstract}
The aim of this contribution is to provide a short introduction to 
 recently investigated models in which 
our accessible universe is a four-dimensional 
submanifold, or brane, embedded in a higher dimensional spacetime and 
ordinary matter is trapped in the brane. I focus here on the gravitational
and cosmological aspects of such models with a single extra-dimension.
\end{abstract}

The traditional view on extra-dimensions is the Kaluza-Klein picture:
 the matter fields live in  compact (usually flat) extra-dimensions, and 
their  Fourier expansion along the extra-coordinates lead to 
an infinite collection of so-called Kaluza-Klein modes, which can 
be interpreted as four-dimensional fields. Their mass spectrum 
is very specific since it is discrete with a mass gap of the order 
of $R^{-1}$, where $R$ is the size of the extra dimensions (common for
 all extra-directions in the simplest cases).
The non-observation of Kaluza-Klein modes in present collider 
experiments therefore provides an upper bound on the size of the 
extra-dimensions: 
\beq
R\lesssim 1 \ ({\rm TeV})^{-1}.
\eeq

Superstring theory requires extra-dimensions to be consistent 
at the quantum level and the Kaluza-Klein compactification 
was invoked to get rid of the superfluous six extra-dimensions, until 
a new picture on extra-dimensions emerged recently,  
such as in the Horava-Witten eleven-dimensional supergravity\cite{hw}. 
In this context, the ordinary matter fields are not 
supposed to be defined everywhere but, in contrast,  are 
assumed to be {\it confined} in  a submanifold, called {\it brane}, 
embedded in 
a higher dimensional space. 

The Horava-Witten model was followed by the radical proposal of 
Arkani-Hamed, Dimopoulos
and Dvali\cite{add}, who, in order to solve the hierarchy problem, suggested 
that we live confined in a three-brane surrounded by $n\geq 2$ 
(flat and compact)
extra-dimensions with a size $R$ as large as the millimeter scale. One could 
then explain the huge value of the Planck mass, with respect to the 
TeV scale, simply as {\it a projection effect}, the relation between the 
fundamental (higher-dimensional) Planck mass $M_{(4+n)}$ and the 
usual four-dimensional Planck mass $M_{(4)}$  being given by
\beq
M_{(4)}^2\sim M_{(4+n)}^{2+n} R^n.
\label{vol_extra}
\eeq
The absence 
 of any observed  deviation from ordinary Newton's law gives 
an  upper bound on the 
compactification radius  \cite{grav_exp}, presently 
of the order of a fraction of millimiter  
($R \lesssim  0.2 \, {\rm mm}$).

Another proposal, even more interesting from the point of view of 
general relativity and cosmology, is due to Randall and Sundrum \cite{rs99b}.
They consider only one extra-dimension but take into account the 
self-gravity of the brane endowed with a tension $\sigma$. 
They moreover assume the presence of a negative cosmological constant $\Lambda$
in the bulk (thus Anti-de Sitter).
Provided the tension of the brane is adjusted so that 
\beq
{\kappa^2\over 6}\sigma={1\over \ell}\equiv \sqrt{-\Lambda/6},
\label{rs}
\eeq
they find a static solution (and mirror-symmetric with respect to the brane) 
of the five-dimensional Einstein equations, 
 described 
by the metric  
\beq
ds^2=e^{-2|y|/\ell}\eta_{\mu\nu} dx^\mu dx^\nu +dy^2,
\eeq
where $\eta_{\mu\nu}$ is the usual Minkowski metric.
Linearized gravity in the brane can be worked out explicitly in this model
 \cite{rs99b,gt99} and one finds that 
the effective gravitational potential reads 
\beq
V(r)={G_{(4)}\over r}\left(1+  {2\ell^2\over 3r^2}\right),
\eeq
where 
 the four-dimensional 
gravitational coupling is given by
\beq
8\pi G_{(4)}=\kappa^2/\ell.
\label{G4}
\eeq
The (approximate) recovery of  the usual Newton's law is due to the 
 presence of 
a zero mode
which is (exponentionally) localized near the brane. The corrections 
to the usual potential, significant only at the AdS scale
$\ell$ and below, arise from a continuum of massive graviton modes.

Let us now turn to cosmology. 
The main motivation for exploring cosmology in models with extra-dimensions 
is that the new effects might  be significant   
 only at very high energies, i.e. in the very early universe. 
Let us thus consider a five-dimensional spacetime with three-dimensional
isotropy and homogeneity, which contains a three-brane corresponding to
our universe. The metric can be written in the form
\beq
ds^2=- n(t,y)^2 dt^2+a(t,y)^2 \delta_{ij}dx^idx^j+dy^2,
\label{metric}
\eeq
with the brane (here spatially flat)
 located at $y=0$.

The energy-momentum tensor can be decomposed into a bulk 
energy-momentum tensor, which is assumed to vanish  here, 
 and a brane energy-momentum tensor, the latter 
being of the form
\beq
 T^A_{\, B}= S^A_{\, B}\delta (y)= \{-\rho_b, p_b, p_b, p_b, 0\}\delta (y),
\eeq
where the delta function expresses the confinement of matter 
in the brane and where  $\rho_b$ and $P_b$ are respectively 
the total energy density and pressure in the brane and depend only
on time.
The presence of the brane induces a jump of the extrinsic curvature 
tensor (defined by $K_{AB}=h_{A}^C\nabla_C n_B$,
where  $n^A$ is the unit vector normal to the brane) related to the 
brane matter content according to  the 
Israel junction conditions 
 \beq
\left[K^A_{\, B}
-K\delta ^A_{\, B}\right]=\kappa^2 S^A_{\, B}.
\label{israel}
\eeq
Still assuming, for simplicity,  mirror (i.e. $Z_2$) symmetry, these 
junction conditions applied to the cosmological metric  
 (\ref{metric}) yield the two conditions
\beq
\left({n'\over n}\right)_{0^+}={\kappa^2\over 6}\left(3p_b+2\rho_b\right),
\qquad
\left({a'\over a}\right)_{0^+}=-{\kappa^2\over 6}\rho_b.
\label{junction}
\eeq
The bulk Einstein equations can be integrated \cite{bdel99}, and the 
substitution of (\ref{junction}) at the location of the brane 
gives the {\it generalized Friedmann equation}

\beq
H_0^2\equiv {\dot a_0^2\over a_0^2}={\kappa^4\over 36}\rho_b^2+{\Lambda\over 6}
+{\C\over a^4},
\label{fried}
\eeq
where $\C$ is an integration constant,
 the subscript `$0$' denoting evaluation at $y=0$.
The  most  remarkable feature of (\ref{fried}) 
 is that the energy density of the brane enters 
quadratically on the right hand side in contrast with the standard 
four-dimensional Friedmann equation where the energy density enters 
linearly. 
As for  
the  energy conservation equation it 
is unchanged in this five-dimensional setup
and still reads
\beq
\dot\rho_b+3H(\rho_b+p_b)=0.
\eeq

In the simplest case where $\Lambda=0$ and $\C=0$, 
 the evolution of the scale factor is given\cite{bdl99} by 
$a\sim t^{1/4}$  (instead 
of  $a\sim t^{1/2}$) for radiation and 
 $a\sim t^{1/3}$  (instead 
of  $a\sim t^{2/3}$) for pressureless 
matter.
Such behaviour  is problematic because it cannot be reconciled 
with the primordial nucleosynthesis scenario, wich 
relies on the competition between  microphysical  reaction rates
 and the expansion rate of the universe. 

Brane cosmology can however be made consistent with nucleosynthesis 
\cite{cosmors} if, following Randall and Sundrum, one introduces a 
negative cosmological constant in the bulk as well as a tension 
in the brane so that   the total energy density 
in the brane, $\rho_b$, splits into 
\beq
\rho_b=\sigma+\rho,
\eeq
where $\rho$ is  the usual cosmological energy density. 
Substituting this decomposition into (\ref{fried}), one gets 
\beq
H^2= \left({\kappa^4\over 36}\sigma^2-{1\over \ell^2}\right)
+{\kappa^4\over 18}\sigma\rho
+{\kappa^4\over 36}\rho^2+{\C\over a^4}.
\eeq
If one fine-tunes the brane tension and the bulk cosmological cosmological 
constant as in (\ref{rs}), 
the first term on the right hand side vanishes.
The second term then becomes the dominant  term if $\rho$ is small enough and
{\it one thus recovers the usual Friedmann equation at low energy}, 
with the identification
$
8\pi G= {\kappa^4}\sigma/6$, 
which is consistent with (\ref{rs}) and 
(\ref{G4}). 

The third term on the right hand side, quadratic in the energy density, 
provides a {\it high-energy correction} to the Friedmann equation 
which becomes significant when the value of the energy density approaches 
the value of the tension $\sigma$ and dominates  at higher 
energy densities. 
Finally, the  radiation-like term, proportional to  $\C$, is 
related to the bulk Weyl tensor. It must be   small enough 
during  nucleosynthesis in order to satisfy the constraints on the 
number of extra light degrees of freedom.

Our model is valid if nucleosynthesis takes place in the low energy 
regime, i.e. $\sigma^{1/4} \gtrsim 1 \ {\rm MeV}$, which implies 
$M \equiv \kappa^{-2/3}\gtrsim 10^4 \ {\rm GeV}$ for the fundamental mass.
However, the requirement to recover ordinary gravity down to scales of the 
submillimeter order gives the tighter constraint
\beq
 M\gtrsim  10^8 \ {\rm GeV}.
\eeq

With this  viable {\it homogeneous} cosmological model, the 
next step is to investigate the  perturbations  from homogeneity, 
to check whether brane cosmology can be made 
 compatible with the current observations and more 
interestingly, whether  it can provide 
  deviations from the standard predictions  
which might be tested in the  future. Both questions are still unanswered
today. 
One can however get  a first flavour of the possible modifications 
by analyzing the equations obtained from the linearized five-dimensional
equations \cite{l00b}. One finds equations similar to the standard 
evolution equations for the perturbations supplemented with 
corrective terms, which are of two types: new terms 
due to the modified   Friedmann equation, which become 
negligible in the low
energy  regime $\rho\ll\sigma$; source terms, 
which come from the bulk perturbations and 
which cannot be determined solely 
from the evolution inside the brane.

\end{document}